\documentclass[12pt]{article} 
\usepackage{amsmath}
\usepackage{a4wide,latexsym,graphicx,epsfig,psfrag}
\usepackage{latexsym,amssymb,amsmath}
\usepackage{cite}
\usepackage{color}

\newcommand{\be}{\begin{equation}}
\newcommand{\ee}{\end{equation}}
\newcommand{\ba}{\begin{eqnarray}}
\newcommand{\ea}{\end{eqnarray}}

\newcommand{\beq}{\begin{equation}}
\newcommand{\eeq}{\end{equation}}

\newcommand{\cL}{{\cal L}}

\newcommand{\bdm}{\begin{displaymath}}
\newcommand{\edm}{\end{displaymath}}
\newcommand{\lgl}{\langle}
\newcommand{\rgl}{\rangle}

\newcommand{\ccdot}{\hskip-0.3ex\cdot\hskip-0.3ex}

\newlength{\dslashwidth}

\def\lsim{\mathrel{\raise.3ex\hbox{$<$\kern-.75em\lower1ex\hbox{$\sim$}}}}
\def\gsim{\mathrel{\raise.3ex\hbox{$>$\kern-.75em\lower1ex\hbox{$\sim$}}}}

\begin{document}
\allowdisplaybreaks

\begin{titlepage}

\begin{flushright}
IFIC/15-27\\
TUM-HEP-998/15 \\
[0.2cm]
\end{flushright}

\vskip1.5cm
\begin{center}
\Large\bf\boldmath On the minimality of the order $p^6$ chiral Lagrangian
\end{center}

\vspace{1cm}
\begin{center}
{\sc P.~Ruiz-Femen\'ia$^{a}$} and  {\sc M.~Zahiri-Abyaneh$^{b}$}\\[5mm]
{\it ${}^a$Physik Department T31,\\
James-Franck-Stra\ss e,
Technische Universit\"at M\"unchen,\\
D--85748 Garching, Germany\\
} \vspace{0.3cm} {\it ${}^b$Instituto de F\'\i sica Corpuscular
(IFIC),
CSIC-Universitat de Val\`encia \\
Apdo. Correos 22085, E-46071 Valencia, Spain}\\[0.3cm]
\end{center}

\vspace{2cm}
\begin{abstract}
\noindent 
A method to find relations between the operators in the 
the mesonic Lagrangian of Chiral Perturbation Theory at order $p^6$
is presented.
The procedure can be used to establish if the basis of operators in the
Lagrangian is minimal. As an example, we apply the method
to the two-flavour case in the absence of scalar and pseudo-scalar sources ($s=p=0$),
and conclude that the minimal Lagrangian 
contains 
27 independent operators.
\end{abstract}
\end{titlepage}

\section{Introduction}

The global chiral symmetry of the QCD Lagrangian for
vanishing quark masses, and its
spontaneous breaking to the diagonal group, 
characterize the strong  interactions among the lightest
hadronic  degrees of freedom --the peudoscalar mesons-- at low energies.
The Nambu-Goldstone nature of these mesons and the mass gap that
separates them from the rest of the hadronic spectrum, allows one
to build an effective field theory (EFT) containing only these
modes, with a perturbative expansion in powers of momenta and
masses. The framework, called Chiral Perturbation Theory (ChPT),
was introduced in its modern form by Weinberg~\cite{Weinberg:1978kz}, and  
Gasser and
Leutwyler~\cite{Gasser:1983yg,Gasser:1984gg}.

At the lowest order, ${\cal O}(p^2)$, the effective ChPT
Lagrangian ${\cal L}_2$ depends only in two low-energy
couplings. One-loop 
contributions built from the lowest-order vertices generate
${\cal O}(p^4)$
divergences that are absorbed by the operators 
of the next-to-leading order $\cL_4$ Lagrangian~\cite{Gasser:1983yg}, 
introducing seven (ten) additional coupling constants
for the two (three) quark flavours case. In the same way, 
taking the computations to the next-to-next-leading order requires
the construction  of the effective Lagrangian at
${\cal O}(p^6)$. This task was first 
performed systematically in Ref.~\cite{Fearing:1994ga}, and
later revisited in~\cite{Bijnens:1999sh}. Through the use 
of partial integration, the equations of motion, Bianchi identities
and the Cayley-Hamilton relations for SU$(n)$ matrices, 
the authors of Ref.~\cite{Bijnens:1999sh} managed to
write down a basis of operators for $\cL_6$ in the even-intrinsic-parity sector
for $n=2$ ($n=3$) light flavours consisting of
90 (53) terms plus 4 (4) contact terms ({\it i.e.} terms not containing the
pseudo-Goldstone fields, which are only needed for renormalization).  
In recent years, an additional relation among the operators
in the basis of~\cite{Bijnens:1999sh} for the $n=2$ case was proven~\cite{Haefeli:2007ty},
where no additional manipulations but those already used  in~\cite{Bijnens:1999sh}
were required. This showed that the derivation of an algorithm to exhaust 
all possible algebraic conditions among the $\cL_6$ operators imposed by 
partial integration, equations of motion, Bianchi identities
and, particularly, Cayley-Hamilton relations, is a nontrivial task.

Therefore, the question about the minimality 
of the ${\cal O}(p^6)$ chiral Lagrangian is proper and, to the best of our knowledge,
remains unanswered.
It is the aim of the present work  to describe a method that provides
necessary conditions for the existence of additional relations
between the operators of the $\cL_6$ Lagrangian, and to show its
application to the two-flavour case 
when massless quarks are considered. 
Our approach does not try to exploit the algebraic conditions mentioned above
(and used in~\cite{Bijnens:1999sh}), but is rather based on the 
analysis of  Green functions built from arbitrary linear combinations of the
operators in the basis.
The 
requirement that the later Green functions must vanish for an arbitrary kinematic configuration
is a necessary condition
for the linear combination to be true at the operator level. From the method one can conclude 
that the basis is minimal when the necessary conditions provide no freedom for the existence of
new relations. On the other hand, if the method allows for new relations, it cannot immediately
answer the question about the minimality of the set, but it has the advantage that it gives
the precise form that the (potential) new relations among the operator must have. With the latter
information at hand, an algebraic
proof of the relation at the operator level shall be greatly simplified. 

The method involves the computation of tree-level Green functions of order $p^6$. Despite being 
tree-level, the large number
of operators in $\cL_6$ and their involved Lorentz structure, containing vertices with up to six derivatives,
produce rather long expressions. The latter can nevertheless be handled easily with the help of
computer tools, and the method lends itself easily to automatization.

The structure of the paper is the following. In
Sec.~\ref{sec:ChPT} we provide the basic ingredients of ChPT
needed for our analysis. The method that searches for further relations 
among the ${\cal O}(p^6)$ operators is described in Sec.~\ref{sec:method},
where details about the calculation of the Green functions which
provide the necessary conditions  are  given through
specific examples.
Its application to the two-flavour case in the chiral limit with 
scalar and pseudo-scalar sources set to zero
is then presented in Sec.~\ref{sec:results}. Finally, we 
give our conclusions in Sec.~\ref {sec:conclusion}.

\section{Chiral perturbation theory}
\label{sec:ChPT}
The effective Lagrangian that implements the spontaneous breaking of the
chiral symmetry SU$(n)_L\times$ SU$(n)_R$ to SU$(n)_V$ in the meson sector
is written as an expansion in powers of derivatives and masses of the pseudo-goldstone
fields~\cite{Weinberg:1978kz,Gasser:1983yg,Gasser:1984gg},
\begin{equation}
\mathcal{L} = {\displaystyle \sum_{n \geqslant 1}} \mathcal{L}_{2n} \,. \label{ChPT}
\end{equation}
The lowest order reads
\begin{equation}\label{Op2ChPT}
\mathcal{L}_2 = \frac{F^2}{4} \langle u_\mu u^\mu+\chi_+ \rangle \,,
\end{equation}
where $F$ is the pion decay constant in the chiral limit and $\langle \dots\rangle$
stands for the trace in flavour space. The chiral tensor $u_\mu$,
\begin{equation}\label{umu}
 u_\mu =
i\left[u^\dagger\left(\partial_\mu-ir_\mu\right)u
                -u\left(\partial_\mu-i\ell_\mu\right)u^\dagger\right]\, ,
\end{equation}
is built from
the Goldstone matrix field
\begin{equation}\label{U}
u= \exp\left(\frac{i}{\sqrt{2}F}\phi\right)\, ,
\end{equation}
and the left and right $n\times n$-dimensional matrix fields in flavour space,
$\ell_\mu=v_\mu - a_\mu$, $r_\mu=v_\mu+a_\mu$,
with $v_\mu$, $a_\mu$ reproducing the couplings of the quarks to the
external vector and axial sources, respectively. On the other hand,
the tensor $\chi_+$ in (\ref{Op2ChPT}) is built from 
$\chi=2B(s+ip)$, with $s$ and $p$ the scalar and pseudo-scalar external
matrix fields and $B$ a low-energy parameter. Quark masses are introduced
in the ChPT meson amplitudes through the scalar matrix $s$. Since  we restrict ourselves in
the specific examples given later to the chiral limit and in addition set $p$ as well as other contributions to $s$ to
zero, we can drop all operators containing the $\chi$ tensor in 
what follows.

In the two flavour-case, which will be used for a specific application of our method,
the matrix $\phi$ collects the pion fields,
\begin{eqnarray}\label{chpt3}
\phi = \left(
                 \begin{array}{cc}
                 \frac{1}{\sqrt{2}}\pi^0& \pi^+\\
                   \pi^- & -\frac{1}{\sqrt{2}}\pi^0
                 \end{array}
               \right)\, .
\end{eqnarray}
The vector and axial external fields are general traceless $2\times2$ matrices,
\begin{eqnarray}\label{chpt5}
v_\mu =\left(
                 \begin{array}{cc}
                   v_{11} & v_{12} \\
                  v_{21} & - v_{11} \\
                 \end{array}
               \right)_{\mu}\
               \qquad\ \textrm{and}\qquad\ a_\mu =\left(
                 \begin{array}{cc}
                   a_{11} & a_{12} \\
                   a_{21}& -a_{11} \\
                 \end{array}
               \right)_{\mu}\, ,
\end{eqnarray}
since we do not confine ourselves to 
the Standard Model vector and axial currents,
but allow for the parametrization of other possible beyond-the-Standard-Model 
currents.

The general structure of the ${\cal O}(p^6)$
ChPT Lagrangian was studied in~\cite{Fearing:1994ga,Bijnens:1999sh}; adopting
the notation of the latter reference, in the $n=2$ case it reads
\begin{equation} \label{Op6ChPT}
{\cal L}_6^{\mathrm{SU}(2)} = \sum_{i=1}^{53} c_i {\cal P}_i + 4\,\, {\rm contact \,\,
terms} ~,
\end{equation}
where ${\cal P}_i$ are the basis elements and $c_i$ are the corresponding low energy constants.
In the massless
limit with scalar and pseudo-scalar sources set to zero, $27+2$ of the $53+4$ operators
in (\ref{Op6ChPT}) remain.  For completeness, we give their explicit form
in the Appendix.

\section{Outline of the method}
\label{sec:method}

We describe next the method used to determine the minimal set of monomials of ${\cal O}(p^6)$
in the
ChPT Lagrangian. It  is based on the trivial identity:
\begin{eqnarray}
\sum_i \alpha_i \,{\cal P}_i &= 0 \ \longrightarrow \ \langle 0|
\,T \, \phi(x_1) \phi(x_2)\dots f_1(y_1) f_2(y_2)\dots \,
\Big( \int d^4x \displaystyle \sum_{i}  \alpha_i \, {\cal P}_i(x) \Big) |0\rangle = 0 \,,
\nonumber\\
 \label{eq:identity}
\end{eqnarray}
that holds if a set of operators ${\cal P}_i$ satisfies a linear
relation with $\alpha_i$ real or complex numbers. The matrix
element on the r.h.s represents a (connected) Green function built
from the insertion of the ${\cal O}(p^6)$ operator
$(\sum_i \alpha_i \, {\cal P}_i)$ and
an arbitrary number of pion  fields ($\phi$), as well as
external field sources ($f_i=v,\, a,\,s,\,p$). For convenience, we shall
work in momentum space, so the Green functions will become
functions of the momenta of the field. 
The relations among operators can produce
different Lagrangians  but must yield the same $S$-matrix elements.
The latter condition allows for the use of the equations of motion at the operator level
for the pion field,
since it can be shown to be equivalent to a redefinition
of the pion field in the generating 
functional~\cite{Arzt:1993gz,Fearing:1994ga,Bijnens:1999sh}.
Consequently, the vanishing of the right hand side
of~(\ref{eq:identity}) is only guaranteed if 
the momenta of the pion fields are taken on the mass shell.
In order to take the on-shell limit we consider the Green functions
to be amputated ones, {\it i.e.} with no propagators attached to the 
external legs. 
For our purposes it is
sufficient to consider the perturbative computation of the Green
function at the leading order in the momentum expansion,
which is ${\cal O}(p^6)$ because the  ${\cal P}_i$ operators  are already
of that order. 

\begin{figure}[t]
\begin{center}
\includegraphics[width=0.8\textwidth]{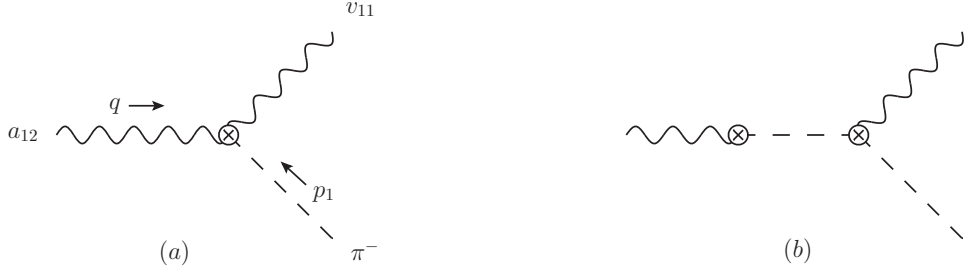}
\caption{$a)$ Local and $b)$ non-local contributions to the $\langle v_{11} a_{12}\pi^-\rangle$ function.}
\label{figavpi}
\end{center}
\end{figure}

The perturbative calculation 
consists of tree-level diagrams, of the form of a contact interaction,
which we shall refer to as ``local" in what follows, as well as
with intermediate pion exchange (``non-local"); see Fig.~\ref{figavpi} for an example.
Local contributions contain an ${\cal P}_i$ operator in the vertex, whereas
non-local contributions have in addition any number of ${\cal O}(p^2)$
vertices, which do not change the chiral order of the amplitude. The Green
functions obtained are rational functions of the momenta,  with a pole structure
given by the propagators present in the diagrams and a numerator
which is a polynomial in the kinematic invariants. If a relation
between operators holds, the Green function must vanish for any
arbitrary momentum configuration of the fields. This requires that
all the coefficients of the terms in the polynomial built from the
kinematic invariants are zero, and conditions for the $\alpha_i$
are thus obtained. 
By requiring that a sufficiently large number of Green functions computed in this way 
with increasing number of fields vanish, we obtain a set of conditions for the numerical
coefficients $\alpha_i$ in (\ref{eq:identity}); when these conditions yield non-zero
solutions, relations between the operators which are fulfilled for all the functions
computed are thus found. One may wish to prove that the relations found hold for
Green functions with an arbitrary number of fields. In that case, the fact that we already
know the precise numerical coefficients in the relation between the operators
simplifies the task of proving it at the operator level using partial integration, equations of motion,
and the Bianchi and Cayley-Hamilton identities. Note also that such a proof may be more a formal matter than
one of practical relevance; processes with 6 mesons legs or involving more than two vector or axial-vector
currents are rather remote experimentally,
so just knowing the relations satisfied among the operators for the 
phenomenologically relevant processes could be enough.

In order to illustrate how the method works let us consider the computation of the Green
function for two specific cases.
The first one corresponds to the matrix element (\ref{eq:identity})
with one external vector ($v_{11}$), one external axial ($a_{12}$) and
one charged pion field ($\pi^-$), which is simple enough to provide explicit formulas.
We shall  refer to the latter with the abridged notation
$\langle v_{11} a_{12}\pi^-\rangle$.

The perturbative computation of this Green function at ${\cal O}(p^6)$ is given by the diagrams
in Fig.~\ref{figavpi}. 
The operators contributing to diagram Fig.~\ref{figavpi}a are ${\cal P}_{44}, {\cal P}_{50},{\cal P}_{51},
{\cal P}_{52}$ and ${\cal P}_{53}$. 
For diagram Fig.~\ref{figavpi}b, operators ${\cal P}_{51},{\cal P}_{52}$ contribute in one of the vertices,
whereas the other vertex corresponds to an ${\cal O}(p^2)$ interaction.
To calculate the amplitude, we take the momenta of the  fields incoming and
use energy-momentum conservation. We thus have two independent momenta,
which we take to be that of the pion, $p_1$, and that of the axial current, $q$. In addition we have the
``polarization" vectors from the 
external fields $v_{11}$ and $a_{12}$, $\epsilon_v$ and $\epsilon_a$ 
respectively.\footnote{The introduction of polarization vectors for the external fields is not
strictly necessary: we could work with the tensor amplitude with  Lorentz indices of the external
sources $\mu,\,\nu$ left open
and require that the coefficients of all tensor structures vanish. The contraction of the 
tensor amplitude with arbitrary vectors $\epsilon_v$, $\epsilon_a$ allows to work with a scalar
function, which simplifies handling the long expressions that are obtained for the amplitudes of 
Green functions with more fields.} Taking into account the on-shell condition for the (massless) pion, $p_1^2=0$,
the amplitude can then be written in terms of seven different Lorentz invariants,
$p_1\cdot q$, $p_1\cdot \epsilon_v$, $p_1\cdot \epsilon_a$,
$q\cdot \epsilon_v$, $q\cdot \epsilon_a$, $\epsilon_v\cdot \epsilon_a$ and $q^2$.
Adding the result from the diagrams with operators ${\cal P}_{i}$
multiplied by corresponding coefficients $\alpha_i$, the perturbative Green function reads
\begin{align}
  \label{amplitude}
G &= \frac{1}{q^2} \Big\{
 4\,( \alpha_ {51} - \alpha_ {53})
 \big[\, \epsilon_v \cdot q \,\epsilon_a\cdot q \,(p_1 \cdot q)^2 - \epsilon_v \cdot p_1\, \epsilon_a \cdot q \,(p_1 \cdot
     q)^2 + q^2\, \epsilon_v \cdot p_1\, \epsilon_a\cdot p_1\, p_1 \cdot q \,\big]
\nonumber\\& \quad\quad \ \
+ (2 \alpha_ {50} - \alpha_ {51} + \alpha_ {52} + \alpha_ {53}) \,q^4 \,\epsilon_v\cdot \epsilon_a \,p_1\cdot q
 + (\alpha_ {51} - \alpha_ {52} \!- \!\alpha_ {53}) \,q^2\, \epsilon_v \cdot q\, \epsilon_a \cdot q \,p_1 \cdot q
\nonumber\\ & \quad\quad \ \
+ (2 \alpha_ {44} - 2 \alpha_ {51} -  \alpha_ {52} +  3 \alpha_ {53})\, 
q^2\, \big[  \, \epsilon_v \cdot p_1 \,\epsilon_a \cdot q \,p_1 \cdot q 
        - q^2\, \epsilon_v\cdot p_1 \,\epsilon_a\cdot p_1 \big]
\nonumber\\& \quad\quad \ \ 
- (2 \alpha_ {44} + 3 \alpha_ {51} - 2 \alpha_ {52} - 2 \alpha_ {53})
    \,q^2  \,\epsilon_v \cdot \epsilon_a\, (p_1\cdot q)^2
\nonumber\\& \quad\quad \ \ 
+ (2 \alpha_ {44}\! - \!\alpha_ {51}\! -\! 2 \alpha_ {52} + 2 \alpha_ {53})
    \, q^2\, \epsilon_v \cdot q\, \epsilon_a \cdot p_1\, p_1\cdot q 
 \nonumber\\ & \quad\quad \ \
 +  2\, \alpha_ {50}\, q^4 \, \big[\, q^2 \,\epsilon_v \cdot \epsilon_a
    - \epsilon_v \cdot q\, \epsilon_a \cdot p_1
    -  \epsilon_v \cdot q \, \epsilon_a \cdot q  \big] \Big\} 
    \,,
\end{align}
up to a global constant factor, and we have also dropped the Dirac delta function with the 
momentum conservation. The $1/q^2$ factor arises from the scalar propagator in diagram
Fig.~\ref{figavpi}a; since we have factored out it globally, 
the resulting polynomial in the numerator is of order $z_i^4$
in the kinematic invariants $z_i\equiv p_1\cdot q, \, p_1\cdot \epsilon_v,\,\dots$.
The requirement that the Green function must vanish if a relation between the 
${\cal O}(p^6)$ operators holds requires that the coefficients of all monomials in the numerator vanish. 
This translates into the following  set of conditions for the  $\alpha_i$:
\begin{eqnarray}
  \label{exresult}
\alpha_ {50} =  \alpha_ {52}= 0 \quad , \quad\alpha_ {51} =  \alpha_ {53}= -2 \alpha_ {44} 
\ .
\end{eqnarray}
The first condition in (\ref{exresult}) implies that no relation involving
operators ${\cal P}_{50}$ and ${\cal P}_{52}$ can be satisfied by the Green function
chosen in this example. Since an operator relation must be true for any Green function 
we can already conclude that the operators 50 and 52 belong to the minimal 
basis of the Lagrangian. The second condition  in (\ref{exresult}) translates into the
relation ${\cal P}_{44}-2 {\cal P}_{51}-2{\cal P}_{53}=0$ being satisfied for this
Green function. By analysing other Green functions we shall conclude in Sec.~\ref{sec:results}
that the later relation is actually part of a larger one involving more terms, that holds
exactly for the operators in ${\cal L}_6^{\rm SU(2)}$.

\begin{figure}[t]
\begin{center}
\includegraphics[width=1\textwidth]{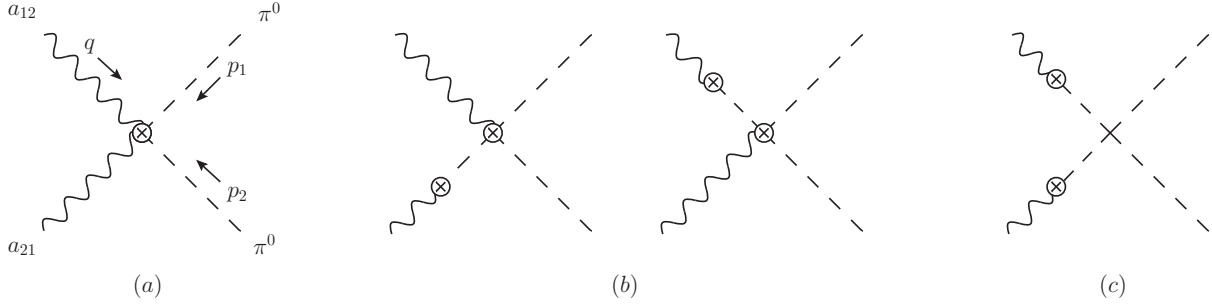}
\caption{$(a)$ Local and $(b),(c)$ non-local contributions to the $\langle a_{12} a_{21}\pi^0\pi^0 \rangle$ function. 
}
\label{figaa2pi}
\end{center}
\end{figure}

Let us now choose a Green function with one field more, for instance
$\langle a_{12} a_{21}\pi^0\pi^0 \rangle$, which includes two axial and two pion
fields. This example shall give us an idea of the increasing complexity related to diagrams with more legs.
Fig.~\ref{figaa2pi} shows the diagrammatic contributions 
to the corresponding Green function. The pure local term,
Fig.~\ref{figaa2pi}a, stems from the operators $1-3,36-44$ and $50-53$. 
The non-local contributions include two different type of diagrams: in Figs.~\ref{figaa2pi}b,
an axial-$3\pi$ vertex from operators $1-3, 36-38$ and $51-53$ of the ${\cal O}(p^6)$  Lagrangian
is combined with the axial-pion vertex from 
${\cal L}_2^{\rm SU(2)}$ \footnote{We note that there is no axial$-\pi$ vertex in ${\cal L}_6^{\rm SU(2)}$.},
whereas in Fig.~\ref{figaa2pi}c, we need the ${\cal O}(p^6)$  4$\pi$ vertices 
from operators ${\cal P}_{1-3}$.
The amplitudes for  $\langle a_{12} a_{21}\pi^0\pi^0 \rangle$ depend on $11$ independent Lorentz invariants,
namely $p_1\cdot p_2,\, q^2,\, p_1\cdot q,\, p_2\cdot q,\, p_1 \cdot\epsilon_{12},\, p_2 \cdot\epsilon_{12},
\, q\cdot \epsilon_{12}, \, p_1\cdot \epsilon_{21},\, p_2 \cdot\epsilon_{21},\,q\cdot \epsilon_{21}$ and
$\epsilon_{12}\cdot \epsilon_{21}$, and we have again considered massless pions.
The number of monomials of order $p^6$ which can be built out
of the kinematic invariants is therefore large, and handling the amplitude in order to find 
out the conditions for the $\alpha_i$ 
requires automatisation. For this task, we have implemented the computation of the tree-level Green
functions at ${\cal O}(p^6)$ and the extraction of the relations for the $\alpha_i$ in a
{\sc Mathematica} code.
In the case at hand, $\langle a_{12} a_{21}\pi^0\pi^0 \rangle$,
one obtains an amplitude with 132 independent monomials in the numerator, whose coefficients yield
the equations for $\alpha_i$: 50 of these equations are non-trivially identical, but only 10 turn out 
to be independent. The solution to this system then provides 10 relations among the coefficients
$\alpha_i$ of the 16 operators that contribute to $\langle a_{12}\, a_{21}\pi^0\pi^0\rangle$:
\begin{gather}
\alpha_{38} = \alpha_{50} =  0  \quad , \quad \alpha_ {1} = -4\alpha_ {2} = \frac{4}{3} \alpha_3 = \alpha_{36} = - \alpha_{37}  \quad , \quad \alpha_{51} = \alpha_{53}
\nonumber\\
 3\alpha_{1}  -2  \alpha_{41} -2 \alpha_{42} + 4 \alpha_{43} -Ê4\alpha_{51} =  0
\nonumber\\[2mm]
\alpha_{1}  + 8 \alpha_{39} -8 \alpha_{40} + 6 \alpha_{41} + 6 \alpha_{42} -12 \alpha_{43} - 8\alpha_{44}=  0
\nonumber\\[2mm]
\alpha_{1}  + 2 \alpha_{39}- 2\alpha_{40}+  \alpha_{41} +  \alpha_{42} - 2 \alpha_{43} -Ê\alpha_{52} =  0
\ .
\label{ex2result} 
\end{gather}
With these conditions, the function $\langle a_{12} a_{21}\pi^0\pi^0 \rangle$
with an insertion of $(\sum_i \alpha_i {\cal P}_i)$ 
thus becomes
\begin{align}
&\Big\langle \alpha_1 \,\Big ( \, {\cal P}_1 -{1\over4}\, {\cal P}_2 + {3\over4}\,{\cal P}_3 
+ {\cal P}_{36} - {\cal P}_{37} - {3\over 4}\, {\cal P}_{40} -  {\cal P}_{41}  - {\cal P}_{42} 
+ {1\over 4}\, {\cal P}_{43} - {\cal P}_{44} + 2{\cal P}_{51} + 2{\cal P}_{53} \Big)
\nonumber \\[2mm]
& \quad + \alpha_{39} \, \Big ( \, {\cal P}_{39} - {\cal P}_{40} -2\,{\cal P}_{41} + {\cal P}_{43} 
- \, {\cal P}_{44} + 2\,{\cal P}_{51} +2\,{\cal P}_{53} \Big)\, \Big\rangle 
= 0  \,, \label{eq:solaapipi}
\end{align}
and vanishes for arbitrary values of $\alpha_1$ and $\alpha_{39}$. The notation $\langle\dots\rangle$ in 
(\ref{eq:solaapipi}) is short for the matrix element of the linear combination together with the axial-vector and pion fields
which defines the Green function. The result (\ref{eq:solaapipi}) implies that the two linear relations
among the ${\cal P}_i$ operators between parenthesis are equal to zero for this particular function.
We can proceed in the 
same way for other Green functions  and require a simultaneous vanishing of
all of them by solving for the $\alpha_i$. The latter is a necessary condition for the existence of
a relation between the ${\cal O}(p^6)$ operators. In the next section we show that the procedure eventually 
allows for just two relations in the SU(2) case without scalar and pseudo-scalar external fields.

\section{SU(2) case with $s=p=0$}
\label{sec:results}
\begin{table*}[t!]
\begin{center}
\renewcommand{\arraystretch}{1.12}
\begin{tabular}{|c|c|c|}
\hline
$\begin{array}{c}
\rm Green \\
\rm function 
\end{array}$
 & ${\cal P}_i$ &
 Operator relations  \\
\hline
\hline
$\langle v v \rangle $ 
& $56$ & $\alpha_{56}\!=\!0$ 
\\\hline
$ \langle v \,2\pi \rangle$
& $51,53$ & ${\cal P}_{51}+{\cal P}_{53} = 0 $ 
\\\hline
$ \langle va \pi \rangle$
  & $44,50\!-\!53 $ & $\alpha_{50}\!=\!\alpha_{52}\!=\! 0$ 
  \\ & & ${\cal P}_{44} - 2 {\cal P}_{51} - 2 {\cal P}_{53} = 0 $ 
  \\\hline
$ \langle v aa \rangle$
  & $ 44,45,$ & $\alpha_{50}\!=\!\alpha_{52}\!=\! 0 $ 
  \\ & $50\!-\!53,55$ & $ 3 {\cal P}_{45} + 8 {\cal P}_{55} = 0 $ 
  \\ & & ${\cal P}_{44} - {\cal P}_{45} - 2 {\cal P}_{51} - 2 {\cal P}_{53} = 0 $ 
\\\hline 
$ \langle 4\pi \rangle$ 
  & $1\!-\!3 $ &   $4 {\cal P}_{1} - {\cal P}_{2} + 3 {\cal P}_{3} = 0 $
\\\hline
$ \langle vv \, 2\pi \rangle$  
  & $ 29\!-\!33,44,$ & $\alpha_{50}\!=\!\alpha_{52} \!=\! 0$ 
  \\ & $50\!-\!53$ & ${\cal P}_{29} - {\cal P}_{30} + 2 {\cal P}_{31} - 4 {\cal P}_{32}
  +  {\cal P}_{33} -  {\cal P}_{44} + 2 {\cal P}_{51} + 2 {\cal P}_{53} = 0 $ 
\\\hline
$ \langle aa\, 2\pi \rangle$
  & $ 1\!-\!3, 36\!-\!44,$ & 
  $\alpha_{38} = \alpha_{50} = \alpha_{52} = 0$ 
  \\ & $50\!-\!53$ & ${\cal P}_{39} - {\cal P}_{40} - 2 {\cal P}_{41} +  {\cal P}_{43}
  - {\cal P}_{44} + 2 {\cal P}_{51} + 2 {\cal P}_{53} = 0 $ 
  \\ & & 
  $\begin{array}{c}
    4{\cal P}_{1} - {\cal P}_{2} + 3 {\cal P}_{3} + 4 {\cal P}_{36}
    -4 {\cal P}_{37} -  4 {\cal P}_{39} + {\cal P}_{40} + 4 {\cal P}_{41} \\ - 4 {\cal P}_{42} -3 {\cal P}_{43}= 0 
  \end{array}$
\\\hline
$ \langle a \, 3\pi \rangle$ 
  & $ 1\!-\!3,36\!-\!38,$  & $\alpha_{38} = 0$
  \\ & $51,53$ & 4${\cal P}_{1} - {\cal P}_{2} + 3 {\cal P}_{3} + 4 {\cal P}_{36}
  -4  {\cal P}_{37} = 0 $ 
  \\ & & $ {\cal P}_{51} + {\cal P}_{53}= 0 $ 
\\\hline
$ \langle vv a \,\pi \rangle$ 
  & $ 29\!-\!33, 44, $  &$\alpha_{50} = \alpha_{52} = 0$ 
  \\ & $45, 50\!-\!53$ &  
  $\begin{array}{c}
    {\cal P}_{29} - {\cal P}_{30} + 2 {\cal P}_{31} - 4  {\cal P}_{32} 
    +  {\cal P}_{33} - {\cal P}_{44} +  {\cal P}_{45} \\ +  2 {\cal P}_{51} + 2 {\cal P}_{53}= 0 
  \end{array}$
\\\hline
$ \langle v \, 4\pi \rangle$
  &  $ 1\!-\!3,27\!-\!28$ & $ \alpha_{38} = 0 $
  \\ & $36\!-\!38,51\!-\!53$ & 4${\cal P}_{1} - {\cal P}_{2} + 3 {\cal P}_{3} - 6 {\cal P}_{27}
  -14  {\cal P}_{28} + 4 {\cal P}_{36} - 4{\cal P}_{37} = 0 $ 
  \\ & &  $2{\cal P}_{27} + 2 {\cal P}_{28} - {\cal P}_{51} -  {\cal P}_{53} = 0$
\\\hline
$\langle va \, 3\pi \rangle$
  &    $ 1\!-\!3, 27\!-\!44,$ & $\alpha_{38} = \alpha_{50} = \alpha_{52} = 0$ 
  \\ & $50\!-\!53$ & $ {\cal P}_{29} + {\cal P}_{39}  = 0 $ 
  \\ & & ${\cal P}_{31} + {\cal P}_{32} +  {\cal P}_{41} +  {\cal P}_{42} = 0 $ 
  \\ & &
  $\begin{array}{c}
    8{\cal P}_{1} -2 {\cal P}_{2} + 6 {\cal P}_{3} - 12 {\cal P}_{27}
    - 28  {\cal P}_{28} + 8 {\cal P}_{36}  - 8 {\cal P}_{37} \\- 8 {\cal P}_{39}
    +2 {\cal P}_{40} + 8 {\cal P}_{41} - 8 {\cal P}_{42} - 6 {\cal P}_{43}= 0 
  \end{array}$
  \\ & & 
  $\begin{array}{c}
    4{\cal P}_{27} + 4 {\cal P}_{28} +  {\cal P}_{30} - 2 {\cal P}_{31}
    +4  {\cal P}_{32} - {\cal P}_{33}  +  {\cal P}_{40}  + 2 {\cal P}_{41}\\
    - {\cal P}_{43} +  {\cal P}_{44} - 2{\cal P}_{51} - 2 {\cal P}_{53}= 0
  \end{array}$ 
\\\hline
$ \langle 6\pi \rangle$ 
  & $1\!-\!3,24\!-\!26 $  &
  $4 {\cal P}_{1} - {\cal P}_{2} + 3 {\cal P}_{3} - 10  {\cal P}_{24}
  + 4  {\cal P}_{25} + 6 {\cal P}_{26} = 0 $ 
\\\hline
\end{tabular}
\end{center}
\caption{Relations among ${\cal O}(p^6)$ operators satisfied for each of the 
Green functions computed. The second column lists the operators that contribute in each case.}
\label{tab1}
\end{table*}

As a proof of concept we show in this section how the method described above applies
to  the two-flavour ChPT Lagrangian in the chiral
limit and without additional external scalar or pseudo-scalar sources ($s=p=0$) but $v_\mu,a_\mu\ne0$.
This simplified  framework does not lack of phenomenological relevance:
it provides a very good approximation to 
the low-energy interaction of the pions in the presence of electroweak currents,
since mass corrections in the $u,d$ quark sector are small and there are no
other contributions to the external sources $s$ and $p$ in the 
Standard Model\footnote{Let us recall that at low energies  the scalar
$q\bar{q}$  interaction
with the Higgs produces terms in the amplitude  suppressed  by $1/m_h^2$.}. 

Within this framework, we have computed the Green functions (\ref{eq:identity}) with the generic field content 
as listed in the first column of Tab.~\ref{tab1}. The notation $\langle va\,3\pi \rangle$, for instance, 
stands for all Green functions involving three pion fields (charged or neutral) and one vector 
and one axial-vector field component, and similarly for the rest of functions. The second column 
indicates which ${\cal O}(p^6)$ operators contribute to the Green functions. The  relations
among the operators satisfied for each function, obtained as in the examples of Sec.~\ref{sec:method} by
solving a system of equations for the coefficients $\alpha_i$,
are then given in the third column. We have not written the equations for the $\alpha_i$ 
for each function
except for the cases where they require some of the $\alpha_i$ to vanish; the condition
$\alpha_i=0$ obtained for a given Green function already implies that the corresponding operator 
${\cal P}_i$ cannot be part of any relation, which is an important information. 
We note that the relations written in Tab.~\ref{tab1}
guarantee that all Green functions with arbitrary charge (or isospin) configuration 
of the pion and external fields vanish. For a given charge (isospin) channel additional 
relations among the operators that contribute can exist, which we do not provide in Tab.~\ref{tab1}. 

The relations satisfied for a set of functions can be obtained by combining
the equations for the coefficients $\alpha_i$ 
from each Green function and looking for a compatible solution. 
From the table one sees that the combination of
Green functions $\langle v v\rangle,\, \langle v aa\rangle,\, 
\langle vv\, 2\pi\rangle,\, \langle aa\,2\pi\rangle$,
$\langle v\, 4\pi\rangle$ and $\langle 6\pi\rangle$ already
involve all the operators in the $\mathrm{SU}(2)$ ChPT Lagrangian with
$s=p=0$. The fact that operators ${\cal P}_{45}$ and ${\cal P}_{55}$ 
only appear in $\langle v aa\rangle$ requires a further Green function
depending on ${\cal P}_{45}$ in order to
fix it completely. That is why the Green function for $\langle v v a\pi\rangle $ is
also computed.
The results for the rest of functions in Tab.~\ref{tab1} is
given for completeness; their computation 
also serves us as a check of the relations found with the minimal
set of functions.

Combining the equations for $\alpha_i$ found for the different Green functions 
we get that all the latter vanish provided
\begin{align}
&\alpha_{38} \, = \alpha_{50} \, = \alpha_{52} \, =
\alpha_{55} \, = \alpha_{56} = 0  \,,
\nonumber\\
& \alpha_1 \,\Big ( \, 8\,{\cal P}_1 -2\, {\cal P}_2 + 6\,{\cal P}_3 -20\,
{\cal P}_{24} +8\,{\cal P}_{25}
                  +12\, {\cal P}_{26} - 16 \,{\cal P}_{28} -3\,{\cal P}_{29} + 3\,{\cal P}_{30} -6  \,{\cal P}_{31} \nonumber\\
                  &\quad\;\;\;+ 12 \,{\cal P}_{32}- 3 \,{\cal P}_{33} +8\,{\cal P}_{36}-8\,{\cal P}_{37}
                  -11\,{\cal P}_{39} + 5\, {\cal P}_{40} + 14\, {\cal P}_{41}  - 8\,{\cal P}_{42} - 9\, {\cal P}_{43}  \nonumber \\
                  &\quad\;\;\;+ 3\,{\cal P}_{44} - 3\,{\cal P}_{45} - 6\,{\cal P}_{51} - 6 \,{\cal P}_{53} \Big)
\nonumber\\
+\; &\alpha_{27} \, \Big ( \, 8\,{\cal P}_{27} + 8\,{\cal P}_{28} -2\,
{\cal P}_{29} + 2\,{\cal P}_{30} -4\, {\cal P}_{31} +8\,{\cal P}_{32}
-2\, {\cal P}_{33} - 2 \,{\cal P}_{39} +2\,{\cal P}_{40} \nonumber \\
&\quad\;\;\;  + 4\,{\cal P}_{41} -2\,{\cal P}_{43} + 2\, {\cal P}_{44} 
-2\, {\cal P}_{45}-4\,{\cal P}_{51} -4\, {\cal P}_{53} \Big)
= 0  \,, \label{eq:solution}
\end{align}
which holds for whatever values of $\alpha_1$ and $\alpha_{27}$,
meaning that the two linear combinations among the operators
${\cal P}_i$ between parenthesis must
vanish independently. The relations obtained can be simplified if
one uses the second linear combination into the first one, which
also allows to compare our result with 
the additional
relation found for the SU(2) case  in~\cite{Haefeli:2007ty},
and with a relation which is known to hold among the 
${\cal O}(p^6)$ operators when  the scalar and pseudo-scalar 
sources are set to zero~\cite{Colangelo:2012ipa}.
In this way we find:
\begin{align}
4\,{\cal P}_{27} + 4\,{\cal P}_{28} -\, {\cal P}_{29} + \,{\cal P}_{30} -2\, {\cal P}_{31}
+4\,{\cal P}_{32}
                  -\,{\cal P}_{33} &-  \,{\cal P}_{39} + \,{\cal P}_{40} + 2\,{\cal P}_{41} \nonumber \\[2mm]
                  - \,{\cal P}_{43} +  \, {\cal P}_{44} - \, {\cal P}_{45}- 2\,{\cal P}_{51} - 2\, {\cal P}_{53}
&= 0 \,, \label{eq:relation1}  \\[3mm]
8\,{\cal P}_1 -2\, {\cal P}_2 + 6\,{\cal P}_3 -20\, {\cal P}_{24} +8\,{\cal P}_{25}
                  +12\, {\cal P}_{26} - 12 \,{\cal P}_{27}& -28\,{\cal P}_{28}+ 8\,{\cal P}_{36}-8\,{\cal P}_{37}  \nonumber\\[2mm]
                  -8\,{\cal P}_{39} +2\, {\cal P}_{40} +8\, {\cal P}_{41}  -8\,{\cal P}_{42}-6\, {\cal P}_{43} 
&= 0  \,, \label{eq:relation2}
\end{align}
that agree with those given 
in~\cite{Colangelo:2012ipa}\footnote{Ref.~\cite{Colangelo:2012ipa} provided
relation~(\ref{eq:relation1}) for a number of flavors $n=3$ using
the SU(3) operator numbering introduced
in~\cite{Bijnens:1999sh}. The corresponding relation for
SU(2) can be obtained by translating into the
SU(2) numbering scheme for the operators, and further
using that the operator ${\cal P}_{52}$ in the two-flavor case
is equal to $-{\cal P}_{50}-{\cal P}_{51}$ ({\it i.e.} to $-{\cal
P}_{27}-{\cal P}_{28}$ in the SU(2) numbering scheme).}
and~\cite{Haefeli:2007ty}, respectively, once the operators depending on 
scalar and pseudo-scalar tensor source $\chi$ are neglected in the latter.
Since relations~(\ref{eq:relation1},\ref{eq:relation2}) were proven algebraically in these references,
they are of course
satisfied for all Green functions of the form (\ref{eq:identity}). We can moreover state that these are
the only two relations between the SU(2) ChPT operators of ${\cal O}(p^6)$ 
in the limit $s=p=0$; otherwise any further relation of the form
$\sum \alpha_i^\prime {\cal P}_i=0$
would have been obtained from the analysis of the functions of Tab.~\ref{tab1}
with our method (let us recall that the vanishing of any Green function with an
insertion of $\sum \alpha_i^\prime {\cal P}_i$ is a necessary condition for the existence of the relation).
We therefore conclude that the set of minimal operators of the SU(2) ChPT Lagrangian
of ${\cal O}(p^6)$ with scalar and pseudo-scalar sources set to zero
reduces from the 27+2 operators initially written down in~\cite{Bijnens:1999sh} to
25+2 (note that the contact terms do not take part in any of the relations above). 
Eqs.~(\ref{eq:relation1},\ref{eq:relation2}) can be used
to drop two of the 27 basis elements of the basis of~\cite{Bijnens:1999sh}.

The application of our method to the general two- and three-flavour cases 
is straightforward. 

For SU(2) including scalar and pseudo-scalar sources, if a similar analysis
does not yield additional relations to that of Ref.~\cite{Haefeli:2007ty},
one would conclude that the basis of ${\cal O}(p^6)$ operators from~\cite{Bijnens:1999sh} 
is minimal up to one term. The case of SU(3) is more involved at the technical level,
since we have to consider an octet of pseudo-goldstone bosons and many more Green
functions can be built. Starting the analysis of Green functions 
with less number of fields, one could expect that the space of solutions for the 
coefficients $\alpha_i$ is either  very much constrained, and eventually no solution
is allowed after computing a few Green functions, or that it  actually 
allows for one (or more) relations 
among the operators. In the former case one could already conclude that  the basis
of ${\cal L}_6^{\rm SU(3)}$ is minimal. In the latter, one may try to check if the 
relations found from the analysis of the simpler Green functions also hold at the level of
the operators ({\it i.e.} for any Green function with an arbitrary number of fields) 
by using the same algebraic manipulations as in~\cite{Bijnens:1999sh}, with 
the great advantage that one would know beforehand the coefficients that the operators
participating in the relation must have. The study of the general two- and three-flavour
cases with the automated tools developed in this work will be the subject 
of future investigation.

\section{Summary}
\label{sec:conclusion}

The large number of low-energy constants in the mesonic  chiral
Lagrangian of order $p^6$  makes their determination by direct comparison with the 
experiment rather difficult. To simplify this task,  one would like to eliminate
possible redundancies by establishing the minimal set of independent
operators in ${\cal L}_6$, that parametrize the rational part of the 
${\cal O}(p^6)$ chiral amplitudes.

We have described in this paper a method to search for additional relations among the
basis  operators
that build the ${\cal O}(p^6)$ SU$(n)$ chiral Lagrangian.
It relies on the computation of tree-level
Green functions with insertions of
the ${\cal L}_6$ operators, which are then required to vanish
for an arbitrary kinematic configuration. 
The method can be used to establish the minimal basis of operators in the
Lagrangian. This has been done in the present work 
for the two-flavour ${\cal O}(p^6)$
Lagrangian  without scalar  and
pseudo-scalar external sources. For this case we have shown that the
original
basis of 27 measurable terms plus 2 contact terms written in~\cite{Bijnens:1999sh}
in the even-intrinsic-parity sector
has 25+2 independent terms, where  the two additional relations between 
operators that emerge from our method
had been already noticed in the literature~\cite{Haefeli:2007ty,Colangelo:2012ipa}. 

As a next step, the method shall be applied to determine 
the minimal basis of operators in the SU(2) case with
general scalar and pseudo-scalar sources, as well as in SU(3). 
Furthermore, one can expect that the method extends naturally
to  other relevant effective actions containing 
a large number of operators, and in particular to the linear and non-linear
effective theories that describe the breaking of the electroweak symmetry.

 \vspace*{0.2cm} \noindent
\subsubsection*{Acknowledgments}
We thank J.~Sanz-Cillero for pointing us to the operator relation
in Ref.~\cite{Colangelo:2012ipa} and for his comments on the draft. 
We also thank G.~Ecker for
useful communication regarding the functionality of the code
Ampcalculator~\cite{ampcalculator}, used for cross-checks, and
J.~Portol\'es for comments on the draft. 
MZA wants to thank A.Pich and J.~Portol\'es for helpful discussions.
This work is partially supported by MEC
(Spain) under grants FPA2007-60323 and FPA2011-23778 and by the
Spanish Consolider-Ingenio 2010 Programme CPAN (CSD2007-00042).

\section{Appendix}

We provide in this appendix the explicit form of the operators in the
${\cal O}(p^6)$ ChPT Lagrangian in the SU(2) case~(\ref{Op6ChPT}) without
scalar and pseudo-scalar sources. The expressions are read off from 
the list given in the appendix C of
Ref.~\cite{Bijnens:1999sh} by discarding terms containing the $\chi$ tensor.

Besides the chiral 
tensors already written in Sec.~\ref{sec:ChPT}, the following 
building blocks are needed to construct the operators in Tab.~\ref{tab:tab2}:
\begin{eqnarray}\label{chpt7}
&& h_{\mu\nu} = \nabla^\mu u_\nu +  \nabla^\nu u_\mu\,,
\nonumber\\[2mm]
&& f_\pm^{\mu\nu} = u F^{\mu\nu}_L u^\dagger \pm u^\dagger
F_R^{\mu\nu}u\,,\quad \nabla_\rho f_\pm^{\mu\nu} \, ,
\end{eqnarray}
with the non-abelian field strength tensor built from  the right and left external fields,
\begin{eqnarray}
&&F_L^{\mu\nu}=\partial^\mu
\ell^\nu-\partial^\nu \ell^\mu -i\left[\ell^\mu,\ell^\nu\right]\,,
\nonumber\\[2mm]
&& F_R^{\mu\nu} =\partial^\mu
r^\nu-\partial^\nu r^\mu -i\left[r^\mu,r^\nu\right]\,
\end{eqnarray}
and the covariant derivative defined as
\begin{equation}\label{chpt8}
\nabla_\mu X = \partial_\mu X + [\Gamma_\mu,X]\, ,
\end{equation}
where
\begin{equation}\label{chpt9}
\Gamma_\mu = \frac{1}{2}\{ u^{\dag} (\partial_\mu - i r_\mu)u + u
(\partial_\mu - i \ell_\mu) u^{\dag} \}\, .
\end{equation}

\begin{table}[t!]
\begin{center}
\renewcommand{\arraystretch}{1.2}
\begin{tabular}{|c|c|c|}
\hline  ${\it i}$ &   ${\cal P}_i$ & Green function \\
\hline \hline
1&  $\lgl u\ccdot u h_{\mu \nu} h^{\mu\nu} \rgl$
	&$ \langle 4\pi \rangle$ \\ 
2&  $\lgl h_{\mu \nu} u_\rho h^{\mu\nu} u^\rho \rgl$ 
	& $\langle 4\pi \rangle$ \\
3& $\lgl h_{\mu \nu} \left(u_\rho h^{\mu\rho} u^\nu + u^\nu h^{\mu \rho} u_\rho \right) \rgl$ 
	& $\langle 4\pi \rangle$ \\
24&   $\lgl (u\ccdot u)^3 \rgl$
	& $ \langle 6\pi \rangle$ \\
25& $\lgl u\ccdot u u_\mu u_\nu u^\mu u^\nu \rgl$  
	& $ \langle 6 \pi \rangle$\\
26 &   $\lgl u_\mu u_\nu u_\rho u^\mu u^\nu u^\rho \rgl$ 
	& $ \langle 6 \pi\rangle$  \\
27&   i $\lgl f_{+ \mu \nu} u_\rho  u^\mu u^\nu u^\rho \rgl$ 
	& $\langle v \, 4 \pi \rangle$ \\
28&   i $\lgl f_{+ \mu \nu} u^\mu u\ccdot u u^\nu \rgl$
	&$ \langle v  \,4 \pi \rangle$ \\
29 &  $\lgl u\ccdot u f_{+ \mu \nu} f_+^{\mu\nu} \rgl$ 
	& $ \langle vv \,2 \pi \rangle$ \\
30&   $\lgl f_{+ \mu \nu} u_\rho f_+^{\mu\nu} u^\rho \rgl$  
	&$ \langle vv \,2\pi \rangle$ \\
31&   $\lgl f_{+ \mu \nu} f_+^{\mu\rho} u^\nu u_\rho \rgl$ 
	&$ \langle vv \,2\pi \rangle$ \\
32&  $\lgl f_{+ \mu \nu} f_+^{\mu\rho} u_\rho u^\nu \rgl$ 
	&$ \langle vv \,2\pi\rangle$ \\
33&   $\lgl f_{+ \mu \nu} \left( u_\rho f_+^{\mu\rho} u^\nu + u^\nu f_+^{\mu \rho}  u_\rho \right) \rgl$ 
	& $\langle vv\, 2 \pi \rangle$ \\
36 &  $\lgl f_{- \mu \nu} \left(h^{\nu\rho} u_\rho u^\mu + u^\mu u_\rho h^{\nu\rho} \right) \rgl$  
	& $ \langle a\, 3\pi\rangle $ \\
37 &   $\lgl f_{- \mu \nu} h^{\nu\rho} \rgl \lgl u^\mu u_\rho \rgl$ 
	& $\langle a \, 3\pi \rangle$ \\
38 &$\lgl f_{- \mu \nu} \left(u^\mu h^{\nu\rho} u_\rho +  u_\rho h^{\nu\rho} u^\mu \right) \rgl$ 
	& $ \langle a \,3\pi \rangle$ \\
39&  $\lgl u\ccdot u f_{- \mu \nu} f_-^{\mu\nu} \rgl$ 
	& $ \langle aa \,2\pi \rangle  $ \\
40&  $\lgl f_{- \mu \nu} u_\rho f_-^{\mu\nu} u^\rho \rgl$
	&$ \langle aa\, 2\pi \rangle$  \\
41&  $\lgl f_{- \mu \nu} f_-^{\mu\rho} u^\nu u_\rho \rgl$
	&$\langle aa\, 2\pi \rangle$  \\
42& $\lgl f_{- \mu \nu} f_-^{\mu\rho} u_\rho u^\nu \rgl$ 
	&$\langle aa\, 2\pi \rangle$  \\
43&   $\lgl f_{- \mu \nu} \left( u_\rho f_-^{\mu\rho} u^\nu + u^\nu f_-^{\mu \rho}  u_\rho \right) \rgl$ 
	& $\langle aa\, 2\pi \rangle$  \\
44 &  i $\lgl f_{+ \mu \nu} [f_-^{ \nu \rho},h^\mu_\rho] \rgl$ 
	& $\langle  va \,\pi \rangle $ \\
45 &   i $\lgl f_{+ \mu \nu} [f_-^{ \nu \rho},f_{-\rho}^\mu] \rgl$ 
	& $\langle v a a \rangle $ \\
50 &  $\lgl \nabla_\rho f_{- \mu \nu} \nabla^\rho f_-^{\mu\nu} \rgl$ 
	&$ \langle a a \rangle$ \\
51 &   i $\lgl \nabla_\rho f_{+ \mu \nu} [h^{\mu\rho}, u^\nu ] \rgl$ 
	& $ \langle v\,2\pi \rangle  $ \\
52 &   i $\lgl \nabla^\mu f_{+ \mu \nu} [f_-^{ \nu \rho}, u_\rho ] \rgl$ 
	& $\langle v a\, \pi \rangle $ \\
53&  i $\lgl \nabla^\mu f_{+ \mu \nu} [h^{\nu\rho}, u_\rho ] \rgl$ 
	& $ \langle v\,2\pi \rangle  $ \\ \hline
contact terms  & &  \\
55&  i $\lgl F_{L \mu \nu} F_L^{ \mu \rho} F_{L \rho}^\nu \rgl + L \rightarrow R $  
	& $ \langle vaa \rangle$  \\
56 &  $\lgl D_\rho F_{L \mu \nu} D^\rho F_L^{ \mu \nu} \rgl + L \rightarrow R$ 
	& $ \langle vv \rangle$  \\
\hline
\end{tabular}
\end{center}
 \caption{\label{tab:tab2} ${\cal O}(p^6)$ operators for $\mathrm{SU}(2)$
 with $s=p=0$, in the basis of~\cite{Bijnens:1999sh}. 
The label in the 
first column refers to the SU(2) numbering scheme used in the latter reference.
The last column indicates the simplest Green function to
which the operator contributes.}
\end{table}


\end{document}